\documentclass[doublecol,sd,amsmath,amssymb,amstext]{epl2}
\usepackage{color}
\usepackage{graphicx}
\usepackage{graphicx,epstopdf}
\usepackage{dcolumn}
\usepackage{bm}
\usepackage{mathrsfs}
\usepackage{amsfonts}
\usepackage{amsmath}
\usepackage{amsthm}

\begin{document}

\title{Dynamics of swarmalators: A pedagogical review}
\author{Gourab Kumar Sar and Dibakar Ghosh \footnote{Email: diba.ghosh@gmail.com}}
\shortauthor{G. K. Sar and D. Ghosh }
\institute{Physics and Applied Mathematics Unit, Indian Statistical Institute, 203 B. T. Road, Kolkata-700108, India}

\date{\today}

\abstract{Swarmalators have emerged as a new paradigm for dynamical collective behavior of multi-agent systems due to the interplay of synchronization and swarming that they inherently incorporate. Their dynamics have been explored with different coupling topologies, interaction functions, external forcing, noise, competitive interactions, and from other important viewpoints. Here we take a systematic approach and review the collective dynamics of swarmalators analytically and/or numerically. Long-term states of position aggregation and phase synchronization are revealed in this perspective with some future problems.}

 {Synchronization; coupled oscillators}
 {Nonlinear dynamical systems}
 {Structures and organization in complex systems}


\maketitle

\section{Introduction}
There is a common saying that ``birds of a feather flock together" which is often metaphorically used to highlight the similar attributes of people who are of same sort. Not only humans, it is found in nature that most of the living organisms have the tendency to stay in a colony and mimic the activities of their neighbors. Flock of birds, school of fish, herd of sheep are some very well-known examples of this very fact \cite{bialek2012statistical, hemelrijk2012schools, sumpter2010collective}. This phenomenon is commonly known as {\it swarming} in the study of multi-agent systems \cite{fetecau2011swarm, reynolds1987flocks}. The examples of swarming are pervasive in coordinated movement of group of animals. To the best of our knowledge, there is no consistent definition of swarming, but swarming systems typically inherit at least one of the two key features: (i) aggregation and (ii) alignment. Similar to swarming, there is another common occurrence which is {\it synchronization}, the examples of which are ubiquitously found in nature and technology \cite{winfree1967biological, kuramoto1975self, pikovsky2001universal}. Chorusing frogs \cite{aihara2008mathematical}, firing neurons \cite{montbrio2015macroscopic}, cardiac dynamics \cite{gois2009analysis}, the dynamics of power grids \cite{motter2013spontaneous} are some instances where synchronization spontaneously occurs. Synchronization usually means self-organization behavior of entities of their internal states in time.

Swarming and synchronization are two very similar, but different phenomena. Swarming is self-organization behavior in space, whereas synchronization can be considered as self-organization behavior in time. In other words, in swarming, the external states of the entities are given core attention, whereas in synchronization, it is the internal states of the units that takes over. Over the past few decades, studies on these two fields have evolved parallelly, though mostly they have remained disconnected. Although in the studies of {\it mobile agents} or {\it moving oscillators}, the effect of agents' motion on the phase dynamics has been contemplated, the reverse scenario has not been considered \cite{vicsek1995novel, frasca2008synchronization, chowdhury2019synchronization, majhi2019emergence, majhi2017synchronization}. Oscillators which can swarm, named {\it swarmalators}, were first introduced by O’Keeffe et al. \cite{o2017oscillators} with the property that the internal and external dynamics are affected by each other and can be identified as a special class of active systems \cite{ramaswamy2010mechanics, marchetti2013hydrodynamics, needleman2017active, shaebani2020computational}. This kind of interplay between sync and swarming is prevalent in nature and has been encountered in different fields, from biology \cite{yang2008cooperation, riedel2005self} and chemistry \cite{zhang2020reconfigurable, bricard2015emergent} to physics \cite{hrabec2018velocity, haltz2021domain} and robotics \cite{barcis2019robots, monaco2020cognitive} etc. Besides this, experimental realizations of the swarmalators systems have been performed in the last decade. {\it Caenorhabditis elegans} are found to synchronize their swimming gait when they are within close proximity \cite{yuan2014gait}. The feedback between the fluid flow generated by catalytic reaction and flexible sheets immersed in a fluid filled chamber enables synchronized motion of the sheets in space and time \cite{manna2021chemical}. A recent experimental study with the nematode {\it Turbatrix aceti} reveals the formation of synchronized collective state whose strength and location are controllable \cite{peshkov2022synchronized}. However, the pioneering work in this direction was carried out by Tanaka et al. \cite{tanaka2007general, iwasa2011juggling} when studying ``chemotactic oscillators" whose movements in space are mediated by a surrounding chemical. 

In this perspective article, we discuss the long-term asymptotic behaviors of the swarmalators under the bidirectional effect of spatial and phase dynamics. We review the theoretical aspects and numerically analyze the long-term states for a lucid knowledge of such systems. By introducing a swarmalator model, we analyze the basic properties, like collision avoidance and minimal inter-particle distance. In the case of global interactions among the constituents, we explore the dynamical states of the model with proper choices of interaction functions and also under the introduction of external forcing. The study is also accomplished with local interactions in the system and in presence of thermal noise. We further introduce attractive-repulsive phase coupling which brings competitive phase interaction and investigate such effect on the model. Finally, for better theoretical understanding of the emerging states, we study the dynamics of swarmalators when they are placed on a $1$D ring.

\section{Model of swarmalators}
The bidirectional interplay between the spatial and phase dynamics of the swarmalators is described by the following pair of equations,
\begin{multline}
	\label{eq.1}
	\dot{\textbf{x}}_{i} = \textbf{v}_{i}+\frac{1}{N} \sum_{\substack{j = 1\\j \neq i}}^{N}\bigg[\text{I}_{\text{att}}(\textbf{x}_{j}-\textbf{x}_{i}) \text{F}_{\text{att}}(\theta_{j}-\theta_{i}) \\- \text{I}_{\text{rep}}(\textbf{x}_{j}-\textbf{x}_{i}) \text{F}_{\text{rep}}(\theta_{j}-\theta_{i})\bigg],
\end{multline}
\begin{equation}
	\label{eq.2}
	\dot{\theta}_{i} = \omega_{i}+\frac{K}{N}\sum_{\substack{j = 1\\j \neq i}}^{N} \text{H}(\theta_{j}-\theta_{i})\text{G}(\textbf{x}_{j}-\textbf{x}_{i}),
\end{equation}
for $i = 1,2, \ldots N$, where $N$ is the number of swarmalators and $\textbf{x}_{i} \in \mathbb{R}^{d}$ ($d = 2$ or $3$), $\theta_{i} \in \mathbb{S}^1$ are the position vector and phase of the $i$th swarmalator, respectively. $\textbf{v}_{i}$ is the self propulsion velocity and $\omega_{i}$ is the natural intrinsic frequency of the $i$th swarmalator. The alignment component of the swarming dynamics in Eq.\ \eqref{eq.1} is reflected through the velocity $\textbf{v}_{i} = (v_0\cos \eta_i, v_0\sin \eta_i)$, where $\eta_i$ is the orientation of the $i$th swarmalator. The orientation of particles (which is another state variable along with $\textbf{x}$ and $\theta$) is not taken into account by the current models of swarmalators and will not be studied in this review. There are spatial attraction and repulsion among the swarmalators, which are regulated by the functions $\text{I}_{att}(\textbf{x})$ and $\text{I}_{rep}(\textbf{x})$, respectively. Swarmalators in nearby phases can attract or repel themselves spatially. The influence of phase similarity of spatial attraction and repulsion are governed by the functions $\text{F}_{att}$ and $\text{F}_{rep}$, respectively. The phase interaction is determined by the function $\text{H}$, where the phase coupling strength is given by $K$. The spatial positions of the swarmalators affect their phase couplings, which is taken into account in the model by the function $\text{G}$. Note that, if there is no effect of phase similarity on spatial attraction or repulsion, i.e., when $\text{F}_{att}(\theta) = 1$ and $\text{F}_{rep}(\theta) = 1$, then Eq.\ \eqref{eq.1} represents pure swarming dynamics. On the other hand, when the effect of spatial position on the phase dynamics is not considered in Eq.\ \eqref{eq.2} (i.e., when $\text{G}(x)=1$), it represents the dynamics of coupled phase oscillators which synchronize beyond a critical phase coupling strength.

\section{Theoretical properties}
The presence of multiple interaction functions in the model makes it difficult to deal analytically. However, analytical progress can be made if we consider a particular instance of the model with the following choices
\begin{equation}
	\text{I}_{att}(\textbf{x})=\frac{\textbf{x}}{|\textbf{x}|^{\alpha}}, \, \text{I}_{rep}(\textbf{x})=\frac{\textbf{x}}{|\textbf{x}|^{\beta}}, \, \text{G}(\textbf{x}) = \frac{1}{|\textbf{x}|^{\gamma}},
	\label{eq.3}
\end{equation}
where $\text{I}_{att}$, $\text{I}_{rep}$, and $\text{G}$ are chosen as power laws with positive exponents $\alpha$, $\beta$, and $\gamma$, respectively, where $|\cdot|$ represents the Euclidean norm. $\alpha$ and $\beta$ play a crucial role in determining the aggregation structure of the system. We want to ensure short-range repulsion and long-range attraction among the swarmalators so that collision among them is avoided and the solution remains bounded \cite{fetecau2011swarm}. For that we assume the exponents $\alpha$ and $\beta$ satisfy the following condition 
\begin{equation}
	1 \le \alpha < \beta.
	\label{eq.6}
\end{equation}
We further assume $\text{F}_{\text{att}}$ and $\text{F}_{\text{rep}}$ to be even and bounded functions of their arguments.
For notational simplicity we set, $\textbf{X} := (\textbf{x}_1,\textbf{x}_2, \ldots,\textbf{x}_{N}),\; \textbf{V} := (\textbf{v}_1,\textbf{v}_2,\ldots,\textbf{v}_{N}),\; \Theta := (\theta_{1},\theta_{2},\ldots,\theta_{N}),\;  W := (\omega_{1},\omega_{2},\ldots,\omega_{N}),\;
\mathcal{N} := \{1,2,\ldots,N\}.$
We define the following functionals
\begin{gather}
	\mathcal{C}_1(\textbf{X},\textbf{V},t) := \sum_{i \in \mathcal{N}} \textbf{x}_i - t \sum_{i \in \mathcal{N}} \textbf{v}_i, \label{eq.12}\\
	\mathcal{C}_2(\Theta,W,t) := \sum_{i \in \mathcal{N}} \theta_i - t \sum_{i \in \mathcal{N}} \omega_i. \label{eq.13}
\end{gather}
Now, taking the sum over Eqs.\ \eqref{eq.1} and \eqref{eq.2} it is elementary to see that $\dfrac{d}{dt}\mathcal{C}_{1,2} = 0$.
So, these two quantities $\mathcal{C}_1$ and $\mathcal{C}_2$ are conserved along the swarmalators dynamics. However, the presence of singular functions like $\frac{1}{|\textbf{x}_{j}-\textbf{x}_{i}|^{\mu}}$ in the model raises the question whether it is well-posed or not. It is imperative to rule out the inter-particle collision, which violates the well-posedness of the model. Inter-particle collision can be avoided if we assume the initial data \((\textbf{X}(0), \Theta(0))\) to be non-collisional i.e., $\min_{1\le i,j \le N} |\mathbf{x}_{i}(0)-\mathbf{x}_{j}(0)| > 0$ for all $i,j \in \mathcal{N}$ and $i \ne j$, along with $\alpha, \beta$ satisfying Eq.\ \eqref{eq.6}. These conditions further guarantee the existence of a minimal inter-particle distance among the swarmalators which eliminates the possibility of unbounded solution in case of swarmalators being asymptotically close. The detailed discussions of these results are found in \cite{ha2019emergent}.

\section{Swarmalator dynamics under global interaction}
We consider a two-dimensional model of swarmalators \cite{o2017oscillators} with specific choices of coupling functions in Eqs. \eqref{eq.1} and \eqref{eq.2}. The spatial functions $\text{I}_{att}$, $\text{I}_{rep}$, and $\text{G}$ are chosen (Eq.\ \eqref{eq.3}) with $\alpha=1, \beta = 2$, and $\gamma=1$ which satisfy Eq.\ \eqref{eq.6}. The influence of phase similarity on spatial attraction is considered by taking the function $\text{F}_{att}(\theta) = 1 + J \cos(\theta)$ while the influence of phase dynamics on the spatial repulsion is not considered (i.e., $\text{F}_{rep}(\theta) = 1$ here). For $J>0$, {\it like attracts like}, i.e., swarmalators in nearby phases attract each other. When $J$ is negative, swarmalators are preferentially attached to the ones in opposite phases. $-1<J<1$ is considered so that $\text{F}_{att}$ is always strictly positive. Phase interaction among the swarmalators is taken in the spirit of Kuramoto model \cite{kuramoto1975self} by taking $\text{H}(\theta) = \sin(\theta)$. Here identical swarmalators are chosen with same velocity $\textbf{v}_i = \textbf{v}$ and frequency $\omega_{i} = \omega$ and by choice of reference frame both of them are set to zero. So, the system now becomes
\begin{equation}
	\label{eq.16}
	\dot{\textbf{x}}_{i} = \frac{1}{N} \sum_{\substack{j = 1\\j \neq i}}^{N}\bigg[\frac{\textbf{x}_{j}-\textbf{x}_{i}}{|\textbf{x}_{j}-\textbf{x}_{i}|} (1+J \cos(\theta_{j}-\theta_{i})) \\- \frac{\textbf{x}_{j}-\textbf{x}_{i}}{|\textbf{x}_{j}-\textbf{x}_{i}|^2}\bigg],
\end{equation}

\begin{equation}
	\label{eq.17}
	\dot{\theta}_{i} = \frac{K}{N}\sum_{\substack{j = 1\\j \neq i}}^{N} \frac{\sin(\theta_{j}-\theta_{i})}{|\textbf{x}_{j}-\textbf{x}_{i}|} .
\end{equation}
\par
\begin{figure}[hpt]
	\centerline{
		\includegraphics[scale = 0.3]{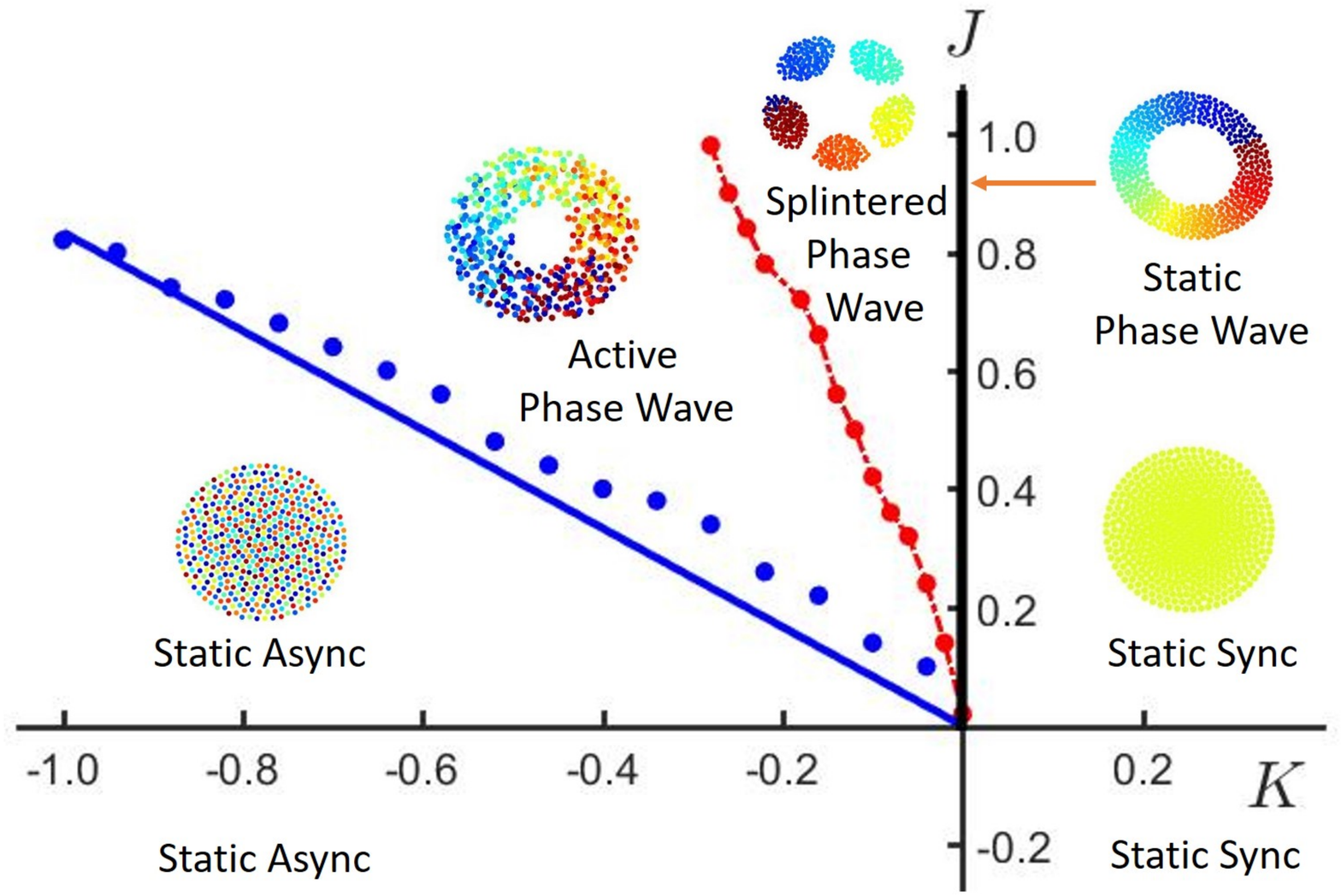}}
	\caption{$K$-$J$ parameter region for emerging collective states. Red dots are the points where $U$ bifurcates from zero to non-zero, found numerically. They are joined with dashed line to separate the regions between splintered phase wave and active phase wave states. Blue dots are numerically calculated points where $S$ bifurcates from non-zero to zero. Blue solid line is the analytical $K=-1.2J$ line where the static async state loses its stability.}
	\label{Fig.1}
\end{figure}

 The spatial and phase couplings among the swarmalators are all-to-all, which means each swarmalator is influenced by the dynamics of other swarmalators present in the system. The two controllable parameters $J$ and $K$ determine the system's long-term dynamical state. Five different asymptotic states are found depending on the values of $J$ and $K$ which is contemplated in Fig.\ \ref{Fig.1}. These states are {\it static sync}, {\it static async}, {\it static phase wave}, {\it splintered phase wave}, and {\it active phase wave}. In the static states, the movement of the swarmalators is ceased and the phases become stationary. There is only one emerging state for $K > 0$ which is the static sync state (see Fig.\ \ref{Fig.1}). The positive value of phase coupling strength minimizes the phase difference among the swarmalators and the phases are completely synchronized in this static state. Three different long-term states are observed for negative values of $K$ which are static async, splintered phase wave, and active phase wave (see Fig.\ \ref{Fig.1}). In the latter two states, swarmalators exhibit movement in the spatial position and their phases also keep evolving and we call them active states. In the absence of phase coupling (when $K=0$), swarmalators' phases are locked to their initial values and for $J>0$ they arrange themselves in the $2$D plane inside an annulus and this state is named static phase wave. Initial positions of the swarmalators are chosen from $[-1,1] \times [-1,1]$ and the initial phases are drawn from $[0,2\pi]$, both uniformly and randomly.
\subsection{Order parameters}
Different steady states emerge between the swarmalators in their spatial and phase dynamics. To investigate the phase coherence, a complex order parameter is defined as $R e^{l {\Phi}} = \frac{1}{N} \sum_{\substack{j = 1}}^{N} e^{l \theta_j}\; (l = \sqrt{-1})$,
where $\Phi$ is the average phase of the swarmalators. $R \in [0,1]$ measures the phase coherence among all the units. In the static sync state, where all the swarmalators' phases are same, the value of $R$ is $1$. In all the other four states, $R$ is strictly less than $1$  as the phases are not synchronized. The correlation between the phases ($\theta_{i}$) and the spatial angles ($\phi_i = \tan^{-1}({y_i}/{x_i})$) is measured with the help of another order parameter, which is defined as
\begin{equation}
	S_{\pm} e^{l \Psi_{\pm}} = \frac{1}{N} \sum_{\substack{j = 1}}^{N} e^{l(\phi_j \pm \theta_j)}.
	\label{eq.19}
\end{equation}
By definition, the magnitude of $S_{\pm}$ varies within $0$ and $1$. These values depend on the choices of initial conditions and for this we define $S = \max(S_+,S_-)$. Now, a nonzero value of $S$ indicates the phases and spatial angles are correlated. In the static phase wave state, the phases are perfectly correlated with the spatial angles, $\theta_{i} = \pm \phi_i + C$ ($\pm$ and $C$ depend on the initial conditions), and $S$ is exactly $1$ here. The correlation decreases when the value of $K$ is reduced from $0$ to negative and it finally reaches the minimum value $0$ in the static async state where swarmalators' phase are distributed uniformly between $0$ to $2\pi$. It is seen that the static async loses its stability beyond a critical phase coupling strength $K_c = -1.2J$ \cite{o2017oscillators} which is highlighted in Fig.\ \ref{Fig.1}. In both the active states (splintered phase wave and active phase wave), the value of $S$ is nonzero, but strictly less than 1. In the splintered phase wave state, swarmalators splinter into clusters and inside each cluster swarmalators move and the phases oscillate around their mean. Swarmalators do not travel from one cluster to another and form disjoint clusters. The number of such clusters and number of swarmalators inside each cluster depend on the initial conditions. Cluster formation vanishes and swarmalators execute regular circular motion both in phase and spatial angle in the active phase wave state. Conveniently, an order parameter $U$ is defined which measures the fraction of swarmalators that execute at least one full circle both in space and phase, after transient period. We define $U = N_{rot}/N$, 
where $N_{rot}$ is the total number of swarmalators executing at least one full circle in space and phase. In Fig.\ \ref{Fig.1}, $K$-$J$ parameter region is demonstrated with the region of occurrence of the five emerging states. Snapshots of the swarmalators' position at $t=1000$ time unit are also included where they are colored according to their phases.


\subsection{Effect of phase similarity both on spatial attraction and repulsion}
\begin{figure}[hpt]
	\centerline{
		\includegraphics[scale = 0.26]{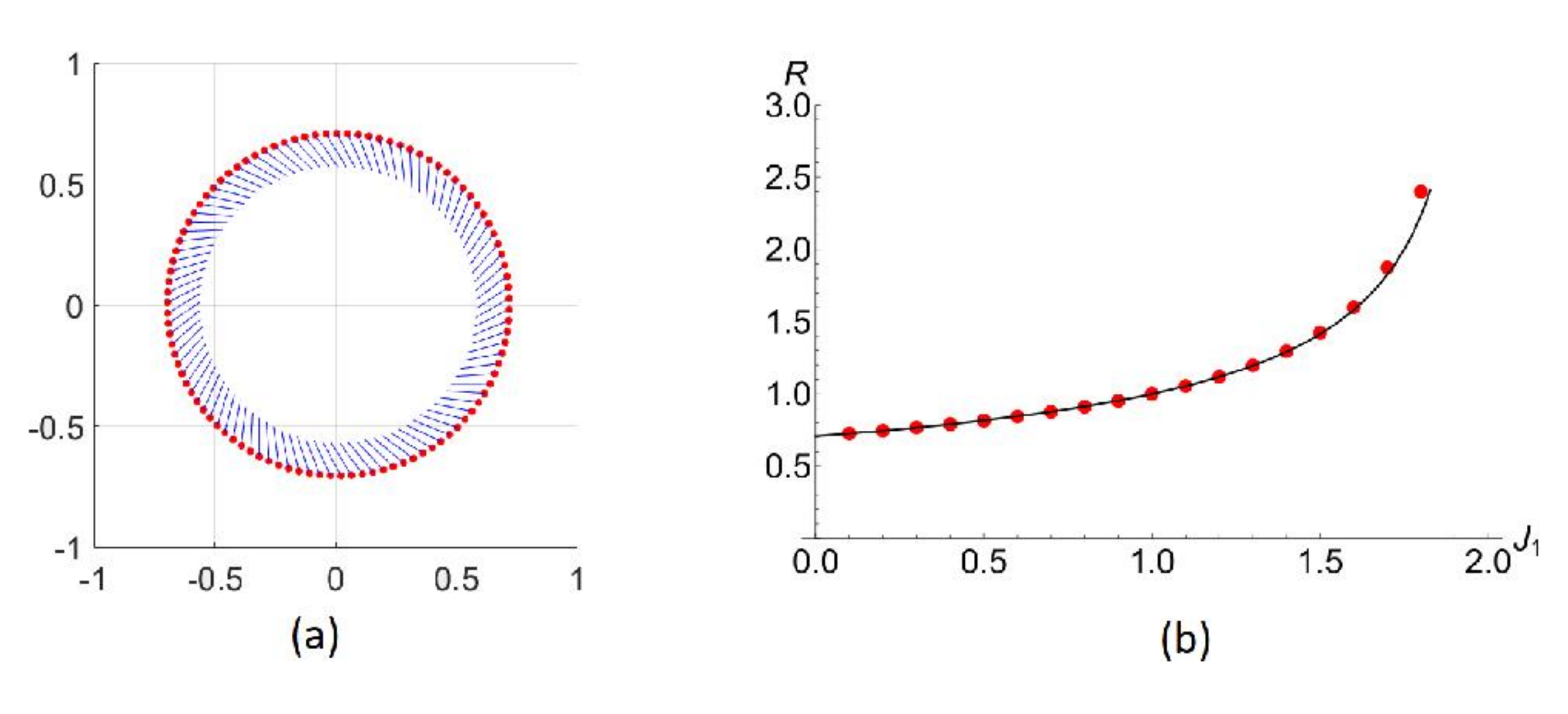}}
	\caption{(a) Scatter plot of ring phase wave state. Here the phase of each swarmalator is represented by a blue ray, and is given by the angle the ray makes with the positive $x$ axis. Ring state is shown for $J_1=0$, $J_2=1$, $K=-0.003$, and $N=100$. (b) Radius of the ring state is plotted against $J_1$. Red dots are numerically calculated values of radius for $J_2=1$ and $N=100$. Black curve is the theoretical one. We refer the reader to ref.\ \cite{o2018ring} for further details.}
	\label{Fig.2}
\end{figure}

 Now, we consider $\text{F}_{att}(\theta) = 1 + J_1\cos(\theta)$ and $\text{F}_{rep}(\theta) = 1 - J_2\cos(\theta)$ to investigate the behavior of swarmalators when phase similarity affects both spatial attraction and repulsion, respectively \cite{o2018ring}. A different set of values for $\alpha$, $\beta$, and $\gamma$ is chosen ($\alpha=0$, $\beta = 2$, and $\gamma=2$). Note that, since Eq.\ \eqref{eq.6} does not hold for these choices of $\alpha, \beta$, and $\gamma$, collision avoidance can not be assured (although numerical simulations show that the particles do not collide as long as $J_2 \le 1$), but these choices simplify the mathematical analysis of the emerging dynamical states provided the model is well-posed. Depending on the system parameters $J_1$, $J_2$, $K$, and $N$, swarmalators display different behavioral dynamics. For certain parameter values, we find a stationary state where swarmalators arrange themselves on a ring centered around the origin. The phases are perfectly correlated with the spatial angles (i.e., $\theta_i = \phi_i + C$, for some constant $C$ depending on initial conditions). Accordingly, this state is named {\it ring phase wave}. The position and phase of the $k$th swarmalator in this state can be expressed as
\begin{gather}
	\textbf{x}_k = R\cos(2\pi k / N) \hat{i} + R\sin(2\pi k/N) \hat{j},
	\label{eq.21}\\ 	\theta_k = 2\pi k/N + C,
	\label{eq.22}
\end{gather}
where $R$ is the radius of the ring, $\hat{i}$ and $\hat{j}$ are unit vectors along $x$ and $y$ directions, and $C$ is a constant determined by the initial conditions. Following the nature of the ring phase wave, it is convenient to move to complex notation where the vector $\textbf{x}_k \equiv (x_k,y_k) \in \mathbb{R}^2$ is identified as a point $z_k = x_k + i y_k$ in the complex plane. The existence and stability analysis involves complex identities which are easier to handle with our aforementioned choice $\alpha=0, \beta=2, \gamma = 2$. We use Eqs.\ \eqref{eq.21} and \eqref{eq.22} and substitute them into the governing equations to find the radius of the ring state, which is $R = \sqrt{(N-1+J_2)/(N(2-J_1))}$. 
So, ring state will exist only in the parameter region $\{J_1<2,\, J_2> 1-N\} \cup \{J_1>2,\, J_2<1-N\}$. In Fig.\ \ref{Fig.2}, we display the organization pattern and radius of this state. The stability analysis of this state is done elaborately in ref.\ \cite{o2018ring}. When $K = 0$, this state loses its stability beyond a critical value of $N$, $N_{max}=8/((2-J_1)(1-J_2))$. Above this value, static phase wave state emerges. For $K<0$, it bifurcates into the splintered phase wave state. Ring phase wave can also be found for the previously chosen values of $\alpha, \beta, \gamma$ ($\alpha=1, \beta=2, \gamma=1$), where the expression of the radius $R$ takes slightly complicated form \cite{Sar_2022}.

\subsection{Swarmalators under external forcing}
Think about a group of fireflies in a forest flashing simultaneously. If a flashing LED is placed in the center of the group, the flashing rhythms of the individuals get affected and synchronize with the LED flash. This kind of external perturbation can introduce new phenomena to the swarmalator system. To study the dynamics of swarmalators under external force, Lizarraga et al.  \cite{lizarraga2020synchronization} introduced periodic forcing which directly affects the phases. This external stimulus is placed at the center of the initial positions of the swarmalators. The phase dynamics under the effect of periodic forcing is now given by
\begin{equation}
	\label{eq.24}
	\dot{\theta}_{i} = \frac{K}{N}\sum_{\substack{j = 1\\j \neq i}}^{N} \frac{\sin(\theta_{j}-\theta_{i})}{|\textbf{x}_{j}-\textbf{x}_{i}|} + F \frac{\cos(\Omega t - \theta_{i})}{|\textbf{x}_{0}-\textbf{x}_{i}|}, 
\end{equation}
where $F$, $\Omega$, and $\textbf{x}_0$ represent the amplitude, frequency, and spatial position of the stimulus, respectively. The spatial dynamics here is same as Eq.\ \eqref{eq.16}. 

\begin{figure}[hpt]
	\centerline{
		\includegraphics[scale = 0.27]{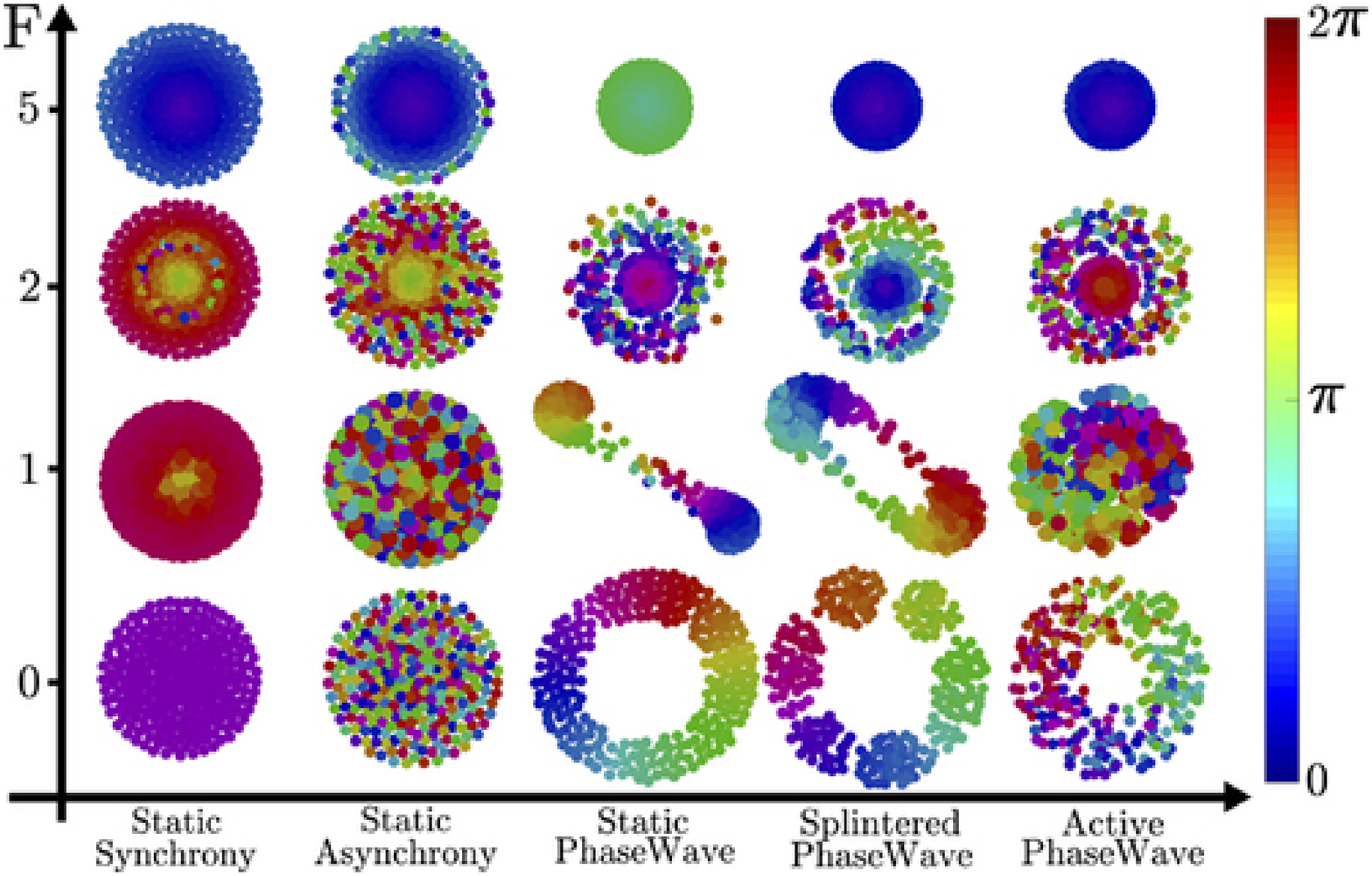}}
	\caption{Transition of states for each of the five non-forced states as a function of force amplitude $F$ for $\Omega = 3\pi/2$. $J=0.1$, $K=1$ for static synchrony, $J=0.1$, $K=-1$ for static asynchrony, $J=1$, $K=0$ for static phase wave, $J=1$, $K=-0.1$ for splintered phase wave, and $J=1$, $K=-0.75$ for active phase wave state are chosen. For further details, see ref.\ \cite{lizarraga2020synchronization}.}
	\label{Fig.3}
\end{figure}

When the amplitude of the force $F$ increases, swarmalators which are near the center of the force start to synchronize with the external frequency $\Omega$. All the five states of non-forced model (static sync, static async, static phase wave, splintered phase wave, and active phase wave) go through phase transition from partial to full synchronization with increasing $F$. Complete synchronization of swarmalators' phases takes place after a critical value of $F$ which is independent of $\Omega$. In the static sync state, the swarmalators near the position of the stimulus get synchronized with the stimulus forming small cluster around it. As the amplitude of the force increases, the cluster also expands until encompassing the whole system (see first column of Fig.\ \ref{Fig.3}). Same phenomena is observed when external forcing is introduced to the static async state only with the exception that swarmalators' phases are now desynchronized in the absence of external force. The dynamics for static phase wave and splintered phase wave state are more complex. When $F=0.5$, swarmalators near the source of the stimulus begin to rotate around it clockwise, whereas those at the boundary show counter-clockwise motion. As the intensity of the force increases to $F=1$, swarmalators divide into two clusters which rotate slowly around the source (see third and fourth columns of Fig.\ \ref{Fig.3}). When $F$ is increased further, the clusters reassemble to synchronize their phases and build small cluster around the source. The size of the cluster grows with increasing $F$ until it covers the whole system. The size of the synchronized state is much smaller here compared to static sync and static async states since the value of $J$ for static phase wave and splintered phase wave are much higher. The transition of the active phase wave state under the stimulus is similar to the splintered phase wave state, but without splitting in groups. These phase transitions are depicted in Fig.\ \ref{Fig.3}.

\section{Dynamics under local interaction}
In the previous section, we have studied the swarmalators where the spatial and phase interactions among them were global. But, in reality most of the real-world multi-agent systems manifest local neighbor interaction among the agents. This motivates us to study the swarmalators where the spatial or phase interaction among the units are restricted to local neighbors. Taking a step in this direction, Lee et al. \cite{lee2021collective} studied the steady state patterns of swarmalators under finite cutoff interaction in the spatial dynamics. Now the swarmalators spatially interact with each other only within a finite range $r$. The spatial dynamics in this scenario is expressed as,
\begin{multline}
	\label{eq.25}
	\dot{\textbf{x}}_{i} = \frac{1}{N_i(r)} \sum_{j \in \Lambda_i(r)}\bigg[(\textbf{x}_{j}-\textbf{x}_{i}) (1+J \cos(\theta_{j}-\theta_{i})) \\- \frac{\textbf{x}_{j}-\textbf{x}_{i}}{|\textbf{x}_{j}-\textbf{x}_{i}|^2}\bigg],
\end{multline}
where $\Lambda_i(r)$ is the set of indices of swarmalators lying within a distance $r$ from the $i$th swarmalator (except itself) and $N_i(r)$ is the cardinality of this set. The phase dynamics is kept same as Eq.\ \eqref{eq.17}. The linear attraction kernel with $\alpha=0$ is considered here to simplify the analysis, although the results remain invariant for $\alpha=1$. When an infinite cutoff is considered, i.e., $r \rightarrow \infty$, all the steady states (static sync, static async, and static phase wave) for global interaction reappear. The system reveals fascinating results when a finite value of $r$ is considered. For $r> D_{\infty}$, we observe multiple identical copies of the steady states, where $D_{\infty}$ is the diameter of the respective steady states in the $r \rightarrow \infty$ limit. The number of such copies depends crucially on the choice of initial conditions. The reason of such occurrences of multiple copies can be justified by the following reason: Once a group of swarmalators assemble in a circular region of radius $r$ where they are separated from others by a minimum distance $r$, they evolve as a separate group independently. Each group follows the identical dynamics for $r \rightarrow \infty$ limit and accordingly, the diameter of each of them is $D_{\infty}$. Consequently, the minimal separation between the groups is $D_{\infty}$. For $r < D_{\infty}$, anomalous version of static sync and static async states are reported where the density of swarmalators are not uniform. Within a certain interval of $r$ ($1<r<1.8$), bar-like patterns emerge from the static phase wave state. If we consider local attractive phase coupling among the swarmalators, the emergence of multiple non-identical static clusters is observed \cite{Sar_2022}.

\section{Competitive phase interaction}
So far we have discussed the results for swarmalators' long-term states where the phase coupling among the units was either positive (attractive) or negative (repulsive). But the coupling characteristic among the units of many systems is often more complex. It can be found in the neuronal networked systems \cite{hopfield1982neural} and in the calling behavior of Japanese tree frogs \cite{aihara2008mathematical}. This has motivated the research community to explore the dynamics of swarmalators in presence of mixed coupling in the system. In ref.\ \cite{hong2021coupling}, the phase coupling strength $K_{ij}$ between the $i$th and the $j$th swarmalators is chosen randomly from a two-peak distribution
\begin{equation}
	h(K_{ij}) = p \delta(K_{ij}-K_a) + (1-p) \delta(K_{ij}- K_r),
\end{equation}
where $K_a$ and $K_r$ are attractive and repulsive coupling strengths, respectively and $p$ is the probability of attractive coupling. The swarmalator dynamics in this case is governed by Eqs.\ \eqref{eq.16} and \eqref{eq.17}, except the fact that now the phase coupling strengths are chosen randomly and with the symmetric property $K_{ij} = K_{ji}$. The ratio $Q =- {K_r}/{K_a}$ ($>0$) can be chosen as a control parameter of the system. Taking annealed approximation of the quenched coupling, it can be shown that the system undergoes phase transition from desynchronized state to the sync state at a critical value $p_c = Q/(1+Q)$. In the incoherent regime $0<p<p_c$, the existence of nonstationary states like splintered phase wave and active phase wave for suitable values of $J$ is found even with positive coupling among swarmalators with $p>0$. To find the possibility of the nonstationary states for $p>0$, we find the mean velocity $\bar{v}$ defined by $\bar{v} = N^{-1} \sum_{i=1}^{N} \sqrt{\dot{x_i}^2 +  \dot{y_i}^2}$, 
where $\text{x}_i = (x_i,y_i)$. The order parameters $S$, $R$, $U$, and $\bar{v}$ are used to distinguish the states and observe the transition. Introducing mixed randomness in the system, several combinations of deformed patterns are encountered which are analogous to the ``chimera state" observed in oscillators with identical frequency \cite{kuramoto2002coexistence, abrams2004chimera}. These deformed patterns are understood with each annealed average of the mixed couplings.

\subsection{Time-varying phase interaction}
\begin{figure}[hpt]
	\centerline{
		\includegraphics[scale = 0.34]{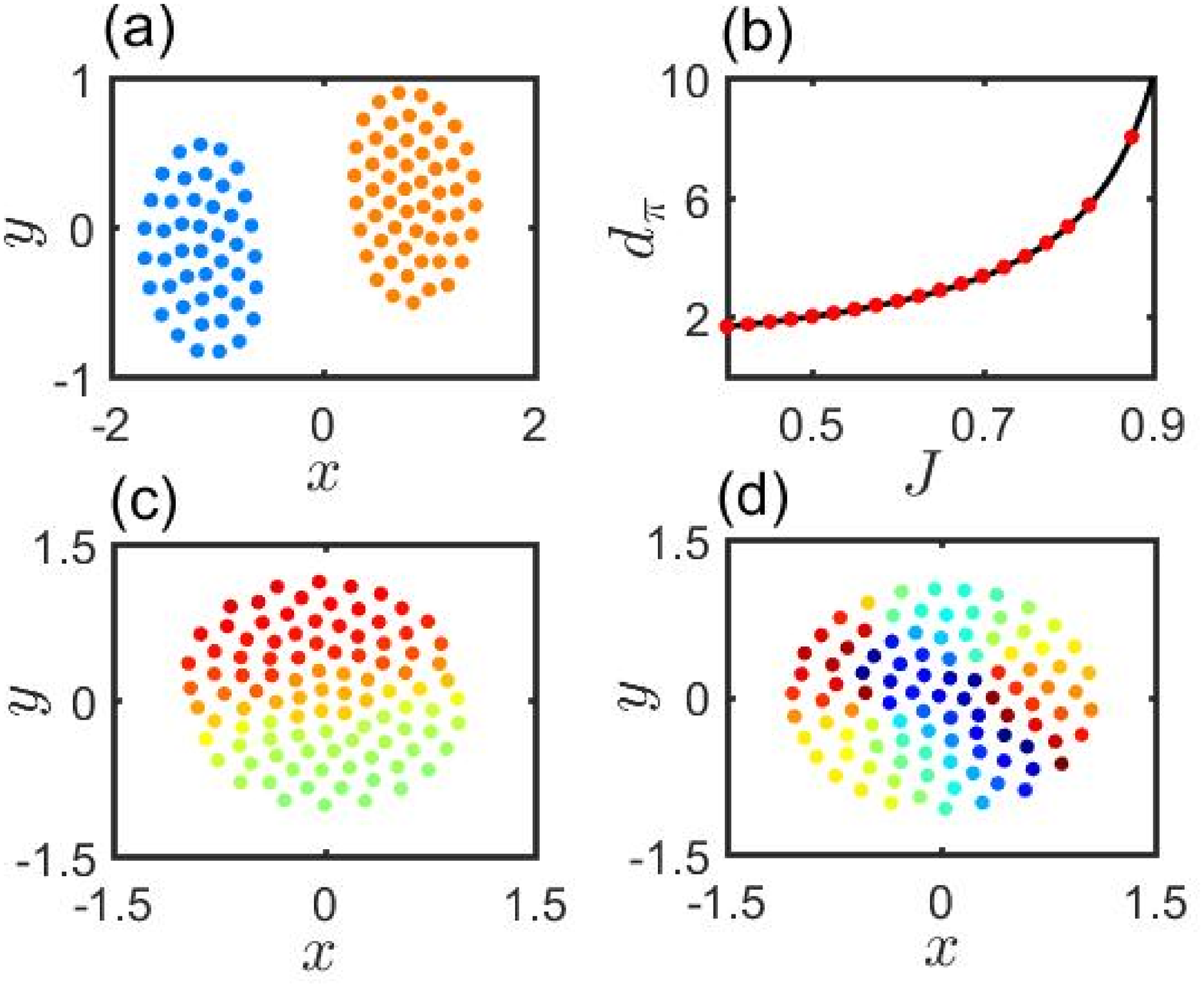}}
	\caption{(a) Snapshot of static $\pi$ state for $(J,r,K_a,K_r)= (0.5,0.8,0.5,-0.5)$. (b) Distance between the center of positions of the clusters in static $\pi$ state. Red dots are numerically calculated values. Black curve is theoretical prediction given by $d_{\pi} = 1/(1-J)$. Snapshots of the mixed phase wave state for (c) $(J,r,K_a,K_r)= (0.1,1.38,0.5,-0.5)$ and (d)$(J,r,K_a,K_r)= (0.1,0.36,0.1,-1.0)$ at $t=1000$ time unit. Simulations are done with $N=100$ swarmalators.}
	\label{Fig.5}
\end{figure}

The studies on swarmalators discussed so far are mainly based on static network formalism, i.e., the coupling scheme among the units does not change over time. It would be interesting to see what happens when we consider time-varying interactions \cite{ghosh2022synchronized} in the system. Sar et al. \cite{Sar_2022} introduced time-varying competitive phase interactions in the swarmalator model and studied their long-term states. Every oscillator moves in the $2$D plane with a uniform circular interaction range, given by the {\it vision radius} $r$. A particular swarmalator is attractively phase coupled with another swarmalator only if it lies inside the interaction range of the former. Otherwise, phase coupling between them is repulsive. Since the swarmalators move in space, the coupling among them changes at every instant of time depending on their spatial positions. This time-varying competitive phase dynamics is given by the equation,
\begin{multline}	
	\label{eq.27}
	\dot{\theta}_{i} =\frac{K_a}{N_{i}(r)}\sum_{\substack{j = 1\\j \neq i}}^{N} \text{A}_{ij}\frac{\sin(\theta_{j}-\theta_{i})}{|\textbf{x}_{j}-\textbf{x}_{i}|} + \\ \frac{K_r}{N-1-N_{i}(r)}\sum_{\substack{j = 1\\j \neq i}}^{N} \text{B}_{ij}\frac{\sin(\theta_{j}-\theta_{i})}{|\textbf{x}_{j}-\textbf{x}_{i}|},
\end{multline}
where $K_a$ and $K_r$ are strengths for attractive and repulsive couplings, respectively and $\text{A}$, $\text{B}$ are the corresponding adjacency matrices.
$N_i(r)$ is same as the one used in Eq.\ \eqref{eq.25}. The spatial dynamics is same as Eq.\ \eqref{eq.16}. In the extreme limits of the vision radius $r$, all the long-term states of the model defined by Eqs.\ \eqref{eq.16} and \eqref{eq.17} are reproduced. By running simulations it is found that, for appropriate values of system parameters $K_a$, $K_r$, $J$, and $r$ the system settles into a stationary state where swarmalators break into two clusters. Within each cluster the phases are totally synchronized, but there is a phase difference of $\pi$ between the clusters. This state is termed as the {\it static $\pi$} state (see Fig.\ \ref{Fig.5}(a)). Analytically it is found that the center of positions of the clusters are always at a distance $d_{\pi} =1/(1-J)$ away from each other which is showed in Fig.\ \ref{Fig.5}(b). For small values of $J$, swarmalators gather themselves in space with the ones in similar phases but do not form clusters. Their phases are neither fully synchronized nor fully desynchronized. The phase difference is small for spatially nearby ones and it increases when the distance increases. In this state, swarmalators show movement in space and this active state is named as {\it mixed phase wave}. In Fig.\ \ref{Fig.5}(c) and (d) snapshots of this state for different parameter values are shown.
\section{Dynamics of swarmalators on a ring}
Although swarmalators show fascinating long-term behavior arising from the bidirectional interplay between spatial and phase dynamics, most of the active states lack analytical support. To address this issue, we can try to reduce the spatial dimension of the system and treat them as positioned on a $1$D ring. This decreases the complexity of the system and allows us to study the states analytically. Now, the model becomes a pair of Kuramoto model which is governed by the equations \cite{o2022collective},
\begin{gather}
	\label{eq.28}
	\dot{x}_i = \frac{J}{N} \sum_{j}^{N} \sin(x_j-x_i) \cos(\theta_j-\theta_i),\\
	\label{eq.29}
	\dot{\theta}_i = \frac{K}{N} \sum_{j}^{N} \sin(\theta_j-\theta_i) \cos(x_j-x_i),
\end{gather}
where $(x_i,\theta_i) \in (\mathbb{S}^1,\mathbb{S}^1)$ denotes the position and phase of the $i$th swarmalator. The parameters $J$ and $K$ control the phase dependent spatial coupling and position dependent phase coupling, respectively. This reduced model given by Eqs.\ \eqref{eq.28} and \eqref{eq.29} exhibits five different dynamical behavior of the swarmalators on a ring. They are named analogously to their $2$D model counterparts as static sync, static $\pi$, static phase wave, static async, and active async following their behavior on the ring. The stability of these observed states are analyzed elaborately in ref.\ \cite{o2022collective}. Similar studies on the ring have been carried out by considering non-identical oscillators \cite{yoon2022sync} and distributed couplings \cite{https://doi.org/10.48550/arxiv.2204.08577}.
\section{Conclusions}
In this perspective, we have made a concise review to explore the dynamical behavior of swarmalator systems. 
 Theoretical analyses have revealed that under certain assumptions such systems possess some conserved quantities and finite-time collision avoidance can be assured. The long-term states of these systems have been analyzed numerically and theoretically under global interaction, addition of external forcing, local interaction, and competitiveness in the phase coupling with both static and time-varying coupling topologies. These studies have enfolded some captivating dynamics. More studies like oscillatory behavior with repulsive short-range interaction \cite{jimenez2020oscillatory}, active phase wave with attractive phase coupling \cite{hong2018active}, dynamics under stochastic coupling and memory \cite{schilcher2021swarmalators}, and mean-field limit \cite{ha2021mean} have enriched this field which we mention here and were unable to discuss in our main text. 
 \par There are plenty to explore about the swarmalators given their complex behavior and interesting self-organizing patterns. Swarmalators with endowed orientation could reveal new collective states where the swarming properties include both aggregation and orientation. The role of the exponents $\alpha,\beta$, and $\gamma$ in determining the asymptotic states is not fully studied yet which demands more attention. A new but impactful direction of research would be to consider non pairwise interaction \cite{majhi2022dynamics} among the units where higher order interactions takes over. The stability properties of active states of the $2$D swarmalator model remain one of the aspects which demand sincere attention. Looking at the difficulty of analyzing the $2$D model, one can ask: Can we model a solvable system for the swarmalators? The $1$D model discussed in the main text is a pivotal step towards this goal which sustains the properties of the $2$D model. The interdependence between the spatial and phase dynamics of the swarmalator systems leads to diverse collective states which emerges as an active field of research.

 

\medskip
\acknowledgments

\end{document}